\documentclass[10pt]{article}
\usepackage{graphicx}
\usepackage{amsmath}
\usepackage{amssymb}
\usepackage{caption2}
\begin{document}
\bibliographystyle{prsty}
\begin{center}
{\large {\bf \sc{  Tetraquark state candidates: $Y(4140)$, $Y(4274)$ and $X(4350)$ }}} \\[2mm]
Zhi-Gang Wang$^{1}$ \footnote{E-mail: zgwang@aliyun.com.  }   and Ye-Fan Tian$^{2} $    \\
$^{1}$ Department of Physics, North China Electric Power University, Baoding 071003, P. R. China \\
$^{2}$ Department of  Physics, University of Science and Technology of China, Hefei 230026, P. R. China
\end{center}

\begin{abstract}
In this article, we tentatively assign the $Y(4140)$, $Y(4274)$ and $X(4350)$ to be the scalar and tensor $cs\bar{c}\bar{s}$ tetraquark states, respectively,
and study them with the QCD sum rules. In the operator product expansion, we take into account   the vacuum condensates up to dimension-10. In calculations,  we use the  formula $\mu=\sqrt{M^2_{X/Y/Z}-(2{\mathbb{M}}_c)^2}$   to determine  the energy scales of the QCD spectral densities.  The numerical results favor assigning the $Y(4140)$ to be the $J^{PC}=2^{++}$   diquark-antidiquark type tetraquark state, and disfavor assigning the $Y(4274)$ and $X(4350)$ to be the $0^{++}$ or $2^{++}$  tetraquark states.
\end{abstract}



\section{Introduction}

In 2009,   the CDF collaboration   observed   a narrow structure
($Y(4140)$) near the $J/\psi\phi$ threshold with statistical
significance in excess of $3.8 \sigma$ in exclusive
$B^+\to J/\psi\phi K^+$ decays produced in $\bar{p} p $ collisions
at $\sqrt{s}=1.96 \,\rm{TeV}$ \cite{CDF0903}. The measured mass and  width   are   $\left(4143.0\pm2.9\pm1.2\right)\,\rm{ MeV}$ and
$\left(11.7^{+8.3}_{-5.0}\pm3.7\right)\, \rm{MeV}$, respectively \cite{CDF0903}. There have been several assignments, such as the molecular state \cite{Y4140-molecule-1,Y4140-molecule-2,Y4140-molecule-3,Y4140-molecule-4,Y4140-molecule-5,Y4140-molecule-6,Y4140-molecule-7,Wang2014-4140}, charmonium hybrid \cite{Y4140-hybrid}, rescattering effect \cite{Y4140-rescattering}, tetraquark state \cite{Y4140-tetraquark}, etc.

Later, the Belle collaboration measured the process $\gamma \gamma \to
\phi J/\psi$ for the $\phi J/\psi$ invariant mass distributions
between the threshold and $5\,\rm{GeV}$, and observed no signal for
 the decay $Y(4140)\to \phi J/\psi$, however, they observed a
narrow peak ($X(4350)$) of $8.8^{+4.2}_{-3.2}$ events with an
significance of $3.2\,\sigma$ \cite{Belle4350}. The measured mass
and width  are $(4350.6^{+4.6}_{-5.1}\pm 0.7)\,\rm{MeV}$ and
$(13.3^{+17.9}_{-9.1}\pm 4.1)\,\rm{MeV}$,  respectively
\cite{Belle4350}.  There also have been several assignments, such as the molecular state \cite{X4350-molecule-1,X4350-molecule-2,X4350-molecule-3,X4350-molecule-4}, conventional charmonium \cite{X4350-charmonium-1,X4350-charmonium-2}, charmonium-molecule mixing state \cite{X4350-mixing}, etc.

 In 2011, the CDF collaboration confirmed the
$Y(4140)$ in the $B^\pm\rightarrow J/\psi\,\phi K^\pm$ decays  with
a  statistical significance greater  than $5\sigma$, the measured mass and width are $\left(4143.4^{+2.9}_{-3.0} \pm0.6
\right)\, \rm{MeV}$ and
$\left(15.3^{+10.4}_{-6.1}\pm2.5\right)\,\rm{MeV}$, respectively
\cite{CDF1101}. Furthermore, the CDF
collaboration  observed an evidence for a second structure ($Y(4274)$) with approximate significance of $3.1\,\sigma$. The
measured mass and width
 are $\left(4274.4^{+8.4}_{-6.7}\pm1.9\right)\,\rm{MeV}$ and
$\left(32.3^{+21.9}_{-15.3}\pm7.6\right)\,\rm{MeV}$, respectively
\cite{CDF1101}. The $Y(4274)$ maybe (or maybe not) a molecular state \cite{Y4274-molecule-1,Y4274-molecule-2,Y4274-molecule-3} or  a $0^{-+}$ tetraquark state \cite{Y4274cscs}.

In 2013, the CMS collaboration   confirmed the $Y(4140)$ in the $J/\psi\phi$  mass spectrum in the $B^\pm \to J/\psi \phi K^\pm$ decays produced in $pp$ collisions at $\sqrt{s} = 7\,\rm{ TeV}$ collected with the CMS detector at the Large Hadron Collider,  and fitted the structure to a $S$-wave relativistic Breit-Wigner line-shape with the statistical significance exceeding $5 \sigma$ \cite{CMS1309}.
Also in 2013,  the D0 collaboration confirmed  the  $Y(4140)$  in the  $B^+ \to J/\psi \phi K^+$ decays in $p\bar{p}$ collisions at $\sqrt{s} = 1.96\,\rm{ TeV}$ collected by the D0 experiment at the Fermilab Tevatron collider with the statistical significance of $3.1\sigma$  \cite{D0-1309}.
The $X(4350)$ and $Y(4274)$ have not been confirmed yet.
 For detailed discussions on this subject, one can consult Ref.\cite{Review-Jpsiphi}.

The S-wave   $J/\psi\phi$ systems  have the quantum numbers $J^{PC}=0^{++}$, $1^{++}$, $2^{++}$, while the P-wave $ J/\psi\phi$ systems have the quantum numbers $0^{-+}$, $1^{-+}$, $2^{-+}$, $3^{-+}$. The $X(4350)$ is observed in the $\gamma\gamma$ fusion, the $J^{PC}=1^{++}$, $1^{-+}$, $3^{-+}$ assignments are excluded due to Yang's Theorem \cite{Review-Jpsiphi}. The possible assignments are $J^{PC}=0^{++}$, $0^{-+}$, $2^{++}$, $2^{-+}$. In  the scenario of tetraquark states, the masses of the $0^{-+}$ and $2^{-+}$ states are much larger than that of the $0^{++}$ and $2^{++}$ states \cite{EFG-2008}. The $Y(4140)$, $X(4350)$ and $Y(4274)$ are observed in the $J/\psi\phi$ invariant mass distribution, if they are tetraquark states,  their quark constituents must be $cs\bar{c}\bar{s}$.   So in this article, we study the masses of the $0^{++}$ and $2^{++}$ $cs\bar{c}\bar{s}$ tetraquark states with the QCD sum rules,  and try to identify the $Y(4140)$, $X(4350)$ and $Y(4274)$.

The article is arranged as follows:  we derive the QCD sum rules for the masses and pole residues of  the scalar and tensor tetraquark states  in section 2; in section 3, we present the numerical results and discussions; section 4 is reserved for our conclusion.

\section{QCD sum rules for  the  scalar  and tensor tetraquark states }
In the following, we write down  the two-point correlation functions $\Pi_{\mu\nu\alpha\beta}(p)$ and $\Pi(p)$ in the QCD sum rules,
\begin{eqnarray}
\Pi_{\mu\nu\alpha\beta}(p)&=&i\int d^4x e^{ip \cdot x} \langle0|T\left\{J_{\mu\nu}(x)J_{\alpha\beta}^{\dagger}(0)\right\}|0\rangle \, , \\
\Pi(p)&=&i\int d^4x e^{ip \cdot x} \langle0|T\left\{J(x)J^{\dagger}(0)\right\}|0\rangle \, ,
\end{eqnarray}
where
\begin{eqnarray}
 J_{\mu\nu}(x)&=&\frac{\epsilon^{ijk}\epsilon^{imn}}{\sqrt{2}}\left\{s^j(x)C\gamma_\mu c^k(x) \bar{s}^m(x)\gamma_\nu C \bar{c}^n(x)+s^j(x)C\gamma_\nu c^k(x)\bar{s}^m(x)\gamma_\mu C \bar{c}^n(x) \right\} \, , \\
 J(x)&=&\epsilon^{ijk}\epsilon^{imn}s^j(x)C\gamma_\mu c^k(x) \bar{s}^m(x)\gamma^\mu C \bar{c}^n(x) \, ,
\end{eqnarray}
the $i$, $j$, $k$, $m$, $n$ are color indexes, the $C$ is the charge conjugation matrix. The
   currents $J_{\mu\nu}(x)$ and $J(x)$ have positive parity and charge conjugation. We  take the   currents $J(x)$ and $J_{\mu\nu}(x)$  to interpolate the scalar  and tensor tetraquark states, respectively.

At the hadronic side, we can insert  a complete set of intermediate hadronic states with
the same quantum numbers as the current operators $J_{\mu\nu}(x)$ and $J(x)$ into the
correlation functions $\Pi_{\mu\nu\alpha\beta}(p)$ and $\Pi(p)$ to obtain the hadronic representation
\cite{SVZ79,Reinders85}. After isolating the ground state
contributions of the scalar  and tensor tetraquark states (denoted by $X$, $Y$ and $Z$), we get the following results,
\begin{eqnarray}
\Pi_{\mu\nu\alpha\beta} (p) &=&\frac{\lambda_{X/Y/ Z}^2}{M_{X/Y/Z}^2-p^2}\left( \frac{\widetilde{g}_{\mu\alpha}\widetilde{g}_{\nu\beta}+\widetilde{g}_{\mu\beta}\widetilde{g}_{\nu\alpha}}{2}-\frac{\widetilde{g}_{\mu\nu}\widetilde{g}_{\alpha\beta}}{3}\right) +\cdots \, \, , \\
\Pi (p) &=&\frac{\lambda_{X/Y/ Z}^2}{M_{X/Y/Z}^2-p^2} +\cdots \, \, ,
\end{eqnarray}
where $\widetilde{g}_{\mu\nu}=g_{\mu\nu}-\frac{p_{\mu}p_{\nu}}{p^2}$, the pole residues  $\lambda_{X/Y/Z}$ are defined by
\begin{eqnarray}
 \langle 0|J_{\mu\nu}(0)|X/Y/Z (p)\rangle &=& \lambda_{X/Y/Z} \, \varepsilon_{\mu\nu}   \, , \nonumber\\
 \langle 0|J (0)|X/Y/Z (p)\rangle &=& \lambda_{X/Y/Z}     \, ,
\end{eqnarray}
the summation of the polarization vector $\varepsilon_{\mu\nu}$
 results in the following formula,
 \begin{eqnarray}
 \sum_{\lambda}\varepsilon^*_{\alpha\beta}(\lambda,p)\varepsilon_{\mu\nu}(\lambda,p)
 &=&\frac{\widetilde{g}_{\alpha\mu}\widetilde{g}_{\beta\nu}+\widetilde{g}_{\alpha\nu}\widetilde{g}_{\beta\mu}}{2}-\frac{\widetilde{g}_{\alpha\beta}\widetilde{g}_{\mu\nu}}{3}\,.
 \end{eqnarray}

 In the following,  we briefly outline  the operator product expansion for the correlation functions $\Pi_{\mu\nu\alpha\beta}(p)$ and $\Pi(p)$ in perturbative QCD.  We contract the $s$ and $c$ quark fields in the correlation functions
$\Pi_{\mu\nu\alpha\beta}(p)$ and $\Pi(p)$ with Wick theorem, and obtain the results:
\begin{eqnarray}
\Pi_{\mu\nu\alpha\beta}(p)&=&\frac{i\epsilon^{ijk}\epsilon^{imn}\epsilon^{i^{\prime}j^{\prime}k^{\prime}}\epsilon^{i^{\prime}m^{\prime}n^{\prime}}}{2}\int d^4x e^{ip \cdot x}   \nonumber\\
&&\left\{{\rm Tr}\left[ \gamma_{\mu}C^{kk^{\prime}}(x)\gamma_{\alpha} CS^{jj^{\prime}T}(x)C\right] {\rm Tr}\left[ \gamma_{\beta} C^{n^{\prime}n}(-x)\gamma_{\nu} C S^{m^{\prime}mT}(-x)C\right] \right. \nonumber\\
&&+{\rm Tr}\left[ \gamma_{\nu} C^{kk^{\prime}}(x)\gamma_{\beta} CS^{jj^{\prime}T}(x)C\right] {\rm Tr}\left[ \gamma_{\alpha} C^{n^{\prime}n}(-x)\gamma_{\mu} C S^{m^{\prime}mT}(-x)C\right] \nonumber\\
&&+{\rm Tr}\left[ \gamma_{\mu} C^{kk^{\prime}}(x) \gamma_{\beta} CS^{jj^{\prime}T}(x)C\right] {\rm Tr}\left[ \gamma_{\alpha} C^{n^{\prime}n}(-x) \gamma_{\nu}C S^{m^{\prime}mT}(-x)C\right] \nonumber\\
 &&\left.+{\rm Tr}\left[ \gamma_{\nu} C^{kk^{\prime}}(x)\gamma_{\alpha} CS^{jj^{\prime}T}(x)C\right] {\rm Tr}\left[ \gamma_{\beta} C^{n^{\prime}n}(-x)\gamma_{\mu} C S^{m^{\prime}mT}(-x)C\right] \right\} \, , \nonumber\\
 \Pi(p)&=&i\epsilon^{ijk}\epsilon^{imn}\epsilon^{i^{\prime}j^{\prime}k^{\prime}}\epsilon^{i^{\prime}m^{\prime}n^{\prime}}\int d^4x e^{ip \cdot x}   \nonumber\\
&&{\rm Tr}\left[ \gamma_{\mu}C^{kk^{\prime}}(x)\gamma_{\alpha} CS^{jj^{\prime}T}(x)C\right] {\rm Tr}\left[ \gamma^{\alpha} C^{n^{\prime}n}(-x)\gamma^{\mu} C S^{m^{\prime}mT}(-x)C\right]   \, ,
\end{eqnarray}
where the   $S_{ij}(x)$   and $C_{ij}(x)$ are the full $s$ and $c$ quark propagators respectively,
 \begin{eqnarray}
S_{ij}(x)&=& \frac{i\delta_{ij}\!\not\!{x}}{ 2\pi^2x^4}
-\frac{\delta_{ij}m_s}{4\pi^2x^2}-\frac{\delta_{ij}\langle
\bar{s}s\rangle}{12} +\frac{i\delta_{ij}\!\not\!{x}m_s
\langle\bar{s}s\rangle}{48}-\frac{\delta_{ij}x^2\langle \bar{s}g_s\sigma Gs\rangle}{192}+\frac{i\delta_{ij}x^2\!\not\!{x} m_s\langle \bar{s}g_s\sigma
 Gs\rangle }{1152}\nonumber\\
&& -\frac{ig_s G^{a}_{\alpha\beta}t^a_{ij}(\!\not\!{x}
\sigma^{\alpha\beta}+\sigma^{\alpha\beta} \!\not\!{x})}{32\pi^2x^2} -\frac{i\delta_{ij}x^2\!\not\!{x}g_s^2\langle \bar{s} s\rangle^2}{7776} -\frac{\delta_{ij}x^4\langle \bar{s}s \rangle\langle g_s^2 GG\rangle}{27648}-\frac{1}{8}\langle\bar{s}_j\sigma^{\mu\nu}s_i \rangle \sigma_{\mu\nu} \nonumber\\
&&   -\frac{1}{4}\langle\bar{s}_j\gamma^{\mu}s_i\rangle \gamma_{\mu }+\cdots \, ,
\end{eqnarray}
\begin{eqnarray}
C_{ij}(x)&=&\frac{i}{(2\pi)^4}\int d^4k e^{-ik \cdot x} \left\{
\frac{\delta_{ij}}{\!\not\!{k}-m_c}
-\frac{g_sG^n_{\alpha\beta}t^n_{ij}}{4}\frac{\sigma^{\alpha\beta}(\!\not\!{k}+m_c)+(\!\not\!{k}+m_c)
\sigma^{\alpha\beta}}{(k^2-m_c^2)^2}\right.\nonumber\\
&&\left. +\frac{g_s D_\alpha G^n_{\beta\lambda}t^n_{ij}(f^{\lambda\beta\alpha}+f^{\lambda\alpha\beta}) }{3(k^2-m_c^2)^4}-\frac{g_s^2 (t^at^b)_{ij} G^a_{\alpha\beta}G^b_{\mu\nu}(f^{\alpha\beta\mu\nu}+f^{\alpha\mu\beta\nu}+f^{\alpha\mu\nu\beta}) }{4(k^2-m_c^2)^5}+\cdots\right\} \, ,\nonumber\\
f^{\lambda\alpha\beta}&=&(\!\not\!{k}+m_c)\gamma^\lambda(\!\not\!{k}+m_c)\gamma^\alpha(\!\not\!{k}+m_c)\gamma^\beta(\!\not\!{k}+m_c)\, ,\nonumber\\
f^{\alpha\beta\mu\nu}&=&(\!\not\!{k}+m_c)\gamma^\alpha(\!\not\!{k}+m_c)\gamma^\beta(\!\not\!{k}+m_c)\gamma^\mu(\!\not\!{k}+m_c)\gamma^\nu(\!\not\!{k}+m_c)\, ,
\end{eqnarray}
and  $t^n=\frac{\lambda^n}{2}$, the $\lambda^n$ is the Gell-Mann matrix,  $D_\alpha=\partial_\alpha-ig_sG^n_\alpha t^n$ \cite{Reinders85}. Then we compute  the integrals both in the coordinate and momentum spaces to obtain the correlation functions $\Pi_{\mu\nu\alpha\beta}(p)$ and $\Pi(p)$ therefore the QCD spectral densities.
In Eq.(10), we retain the terms $\langle\bar{s}_j\sigma_{\mu\nu}s_i \rangle$ and $\langle\bar{s}_j\gamma_{\mu}s_i\rangle$ originate from the Fierz re-arrangement of the $\langle s_i \bar{s}_j\rangle$ to  absorb the gluons  emitted from the heavy quark lines   to extract the mixed condensate $\langle\bar{s}g_s\sigma G s\rangle$ and four-quark condensates  $g_s^2\langle\bar{s}s\rangle^2$, respectively.

 Finally  we can take the
quark-hadron duality below the continuum thresholds  $s_0$ and perform Borel transform  with respect to
the variable $P^2=-p^2$ to obtain  the QCD sum rules:
\begin{eqnarray}
\lambda^2_{X/Y/Z}\, \exp\left(-\frac{M^2_{X/Y/Z}}{T^2}\right)= \int_{4m_c^2}^{s_0} ds\, \rho(s) \, \exp\left(-\frac{s}{T^2}\right) \, ,
\end{eqnarray}
where
\begin{eqnarray}
\rho(s)&=&\rho_{0}(s)+\rho_{3}(s) +\rho_{4}(s)+\rho_{5}(s)+\rho_{6}(s)+\rho_{7}(s) +\rho_{8}(s)+\rho_{10}(s)\, ,
\end{eqnarray}
the explicit expressions of the $\rho_i(s)$ are given in the appendix.

 We differentiate   Eq.(12) with respect to  $\frac{1}{T^2}$, then eliminate the
 pole residues $\lambda_{X/Y/Z}$, and  obtain the QCD sum rules for
 the masses of the scalar   and tensor tetraquark states,
 \begin{eqnarray}
 M^2_{X/Y/Z}= \frac{\int_{4m_c^2}^{s_0} ds\frac{d}{d \left(-1/T^2\right)}\rho(s)\exp\left(-\frac{s}{T^2}\right)}{\int_{4m_c^2}^{s_0} ds \rho(s)\exp\left(-\frac{s}{T^2}\right)}\, .
\end{eqnarray}

\section{Numerical results and discussions}
The vacuum condensates are taken to be the standard values
$\langle\bar{q}q \rangle=-(0.24\pm 0.01\, \rm{GeV})^3$, $\langle\bar{s}s \rangle=(0.8\pm0.1)\langle\bar{q}q \rangle$,
$\langle\bar{s}g_s\sigma G s \rangle=m_0^2\langle \bar{s}s \rangle$,
$m_0^2=(0.8 \pm 0.1)\,\rm{GeV}^2$, $\langle \frac{\alpha_s
GG}{\pi}\rangle=(0.33\,\rm{GeV})^4 $    at the energy scale  $\mu=1\, \rm{GeV}$
\cite{SVZ79,Reinders85,Ioffe2005-1,Ioffe2005-2}.
The quark condensates and mixed quark condensates evolve with the   renormalization group equation,
$\langle\bar{q}q \rangle(\mu)=\langle\bar{q}q \rangle(Q)\left[\frac{\alpha_{s}(Q)}{\alpha_{s}(\mu)}\right]^{\frac{4}{9}}$,
$\langle\bar{s}s \rangle(\mu)=\langle\bar{s}s \rangle(Q)\left[\frac{\alpha_{s}(Q)}{\alpha_{s}(\mu)}\right]^{\frac{4}{9}}$,
  $\langle\bar{s}g_s \sigma Gs \rangle(\mu)=\langle\bar{s}g_s \sigma Gs \rangle(Q)\left[\frac{\alpha_{s}(Q)}{\alpha_{s}(\mu)}\right]^{\frac{2}{27}}$, we take into account the energy scale dependence.

In the article, we take the $\overline{MS}$ masses $m_{c}(m_c)=(1.275\pm0.025)\,\rm{GeV}$  and $m_s(\mu=2\,\rm{GeV})=(0.095\pm0.005)\,\rm{GeV}$
 from the Particle Data Group \cite{PDG}, and take into account
the energy-scale dependence of  the $\overline{MS}$ masses from the renormalization group equation,
\begin{eqnarray}
m_s(\mu)&=&m_s({\rm 2GeV} )\left[\frac{\alpha_{s}(\mu)}{\alpha_{s}({\rm 2GeV})}\right]^{\frac{4}{9}} \, ,\nonumber\\
m_c(\mu)&=&m_c(m_c)\left[\frac{\alpha_{s}(\mu)}{\alpha_{s}(m_c)}\right]^{\frac{12}{25}} \, ,\nonumber\\
\alpha_s(\mu)&=&\frac{1}{b_0t}\left[1-\frac{b_1}{b_0^2}\frac{\log t}{t} +\frac{b_1^2(\log^2{t}-\log{t}-1)+b_0b_2}{b_0^4t^2}\right]\, ,
\end{eqnarray}
  where $t=\log \frac{\mu^2}{\Lambda^2}$, $b_0=\frac{33-2n_f}{12\pi}$, $b_1=\frac{153-19n_f}{24\pi^2}$, $b_2=\frac{2857-\frac{5033}{9}n_f+\frac{325}{27}n_f^2}{128\pi^3}$,  $\Lambda=213\,\rm{MeV}$, $296\,\rm{MeV}$  and  $339\,\rm{MeV}$ for the flavors  $n_f=5$, $4$ and $3$, respectively  \cite{PDG}.

In Refs.\cite{Wang2014-4140,WangHuangTao-1,WangHuangTao-2,WangHuangTao-3,Wang4430,Wang-Cu-Cu,WangHuang-molecule}, we study the acceptable energy scales of the QCD spectral densities for the hidden  charmed (bottom) tetraquark states and molecular states in the QCD sum rules  in details for the first time,  and suggest a  formula,
\begin{eqnarray}
\mu&=&\sqrt{M^2_{X/Y/Z}-(2{\mathbb{M}}_Q)^2} \, ,
 \end{eqnarray}
with the effective $Q$-quark masses ${\mathbb{M}}_Q$ to determine  the energy scales of the QCD spectral densities.
In Refs.\cite{WangHuangTao-1,WangHuangTao-2,WangHuangTao-3,Wang4430,Wang-Cu-Cu}, we focus on the scenario of tetraquark states, study the diquark-antidiquark type scalar, vector, axial-vector, tensor hidden charmed tetraquark states and
axial-vector hidden bottom tetraquark states systematically  with the QCD sum rules,
and try to make possible   assignments of the $X(3872)$,
$Z_c(3900)$, $Z_c(3885)$, $Z_c(4020)$, $Z_c(4025)$, $Z(4050)$, $Z(4250)$, $Y(4360)$, $Z(4430)$, $Y(4630)$, $Y(4660)$, $Z_b(10610)$  and $Z_b(10650)$.
In the operator product expansion, we calculate the  vacuum condensates up to dimension-10, just like in the present case;
 the energy scale  formula  works very well.

In the conventional QCD sum rules \cite{SVZ79,Reinders85}, we
  usually take  the energy gap between the ground
states and the first radial excited states to be $(0.4-0.6)\,\rm{GeV}$. Such relation survives in the tetraquark sector,
 for example,
the  $Z(4430)$ is tentatively assigned to be the first radial excitation of the $Z_c(3900)$ according to the
analogous decays,
$Z_c(3900)^\pm\to J/\psi\pi^\pm$, $Z(4430)^\pm\to\psi^\prime\pi^\pm$,
and the mass differences $M_{Z(4430)}-M_{Z_c(3900)}=576\,\rm{MeV}$, $M_{\psi^\prime}-M_{J/\psi}=589\,\rm{MeV}$ \cite{Wang4430,Maiani-2014,Nielsen-1401}.

Firstly, we take the $Y(4140)$, $Y(4274)$ and $X(4350)$ as the scalar and tensor $cs\bar{c}\bar{s}$ tetraquark states, respectively, and choose the continuum threshold parameters
as $s^0_{Y(4140)}=(4.70\,\rm{GeV})^2$,  $s^0_{Y(4274)}=(4.80\,\rm{GeV})^2$ and $s^0_{X(4350)}=(4.85\,\rm{GeV})^2$.
In Fig.1,
 the masses of the scalar  and tensor  tetraquark states  are plotted  with variations of the  Borel parameters $T^2$ and energy scales $\mu$. From the figure, we can see that the masses decrease monotonously with increase of the energy scales, and we can also obtain the allowed energy scales to reproduce the experimental values of the masses.

 In Table 1, we denote the allowed energy scales which can reproduce the experimental values of the masses  as $\mu_A$, and denote the resulting energy scales from the energy scale formula as $\mu_T$.
 From the table, we can see that the $\mu_A$ and $\mu_T$ are compatible only in the case of the $Y(4140)$ with the assignment $J^{PC}=2^{++}$.

Now, we assume the $Y(4140)$ to be the tensor tetraquark state, take the continuum threshold parameter as $s^0_{Y(4140)}=(4.7\pm 0.1)^2\,\rm{GeV}^2$ and the energy scale as $\mu=2.0\,\rm{GeV}$ to search for
the Borel parameter $T^2$ to satisfy the
two criteria (pole dominance and convergence of the operator product
expansion) of the QCD sum rules. Furthermore, we study the scalar tetraquark state in the same way, i.e.
we search for the optimal Borel parameter $T^2$ and threshold
parameter $s_0$ to satisfy the
two criteria   of the QCD sum rules and the energy scale formula of the QCD spectral densities.
The resulting Borel parameters, continuum threshold parameters and the pole contributions are shown explicitly in Table 2.

In Fig.2,  we plot the contributions of different terms in the
operator product expansion   with variations of the Borel parameters  $T^2$ for the threshold parameters $s^0_{J=2}=(4.7\,\rm{GeV})^2$  and $s^0_{J=0}=(4.5\,\rm{GeV})^2$, respectively.  In the Borel windows,  the $D_0$, $D_3$ and $D_5$
play an important  role,   the $D_6$  and $D_{8}$ play a minor important role, while  the $D_4$, $D_7$ and $D_{10}$ are tiny, where the $D_i$ denote the contributions of the vacuum condensates of dimensions $D=i$. The operator product expansion is well convergent. It is obvious that the two criteria  of the QCD sum rules are fully satisfied, so we expect to make reasonable predictions.

We take into account all uncertainties of the input parameters,
and obtain the values of the masses and pole residues of
 the   scalar and tensor tetraquark states, which are  shown explicitly in Figs.3-4 and Table 2.
The prediction   $M_{J=2} =\left(4.13^{+0.08}_{-0.08}\right)\,\rm{GeV}$ is consistent with the experimental value $M_{Y(4140)}=(4143.0\pm 2.9\pm1.2)\,\rm{MeV}$ \cite{PDG}.   The present predictions favor assigning the $Y(4140)$   to be the $J^{PC}=2^{++}$   diquark-antidiquark type tetraquark states, and disfavor assigning the $Y(4274)$ and $X(4350)$   to be the $J^{PC}=0^{++}$ or $2^{++}$   diquark-antidiquark type tetraquark states.
At the present time, there is no experimental candidate for the scalar $cs\bar{c}\bar{s}$ tetraquark state,
 we can search for the scalar  tetraquark state  at the BESIII, LHCb and Belle-II in the futures.

 Recently, Mo et al study  the $X(4350)$ as a   $ c s\bar{c}\bar{s}$ tetraquark state with the assignment $J^{PC}=1^{-+}$ using the QCD sum rules, and obtain the mass $M_{J=1}=(4.82\pm 0.19)\,\rm{GeV}$, which is not compatible with the $X(4350)$   as a $1^{-+}$ tetraquark state \cite{X4350-Huang}.  So the $X(4350)$ is unlikely to be a $ c s\bar{c}\bar{s}$ tetraquark state. Furthermore, the $X(4350)$ and $Y(4274)$ are still need confirmation.

\begin{figure}
\centering
\includegraphics[totalheight=6cm,width=7cm]{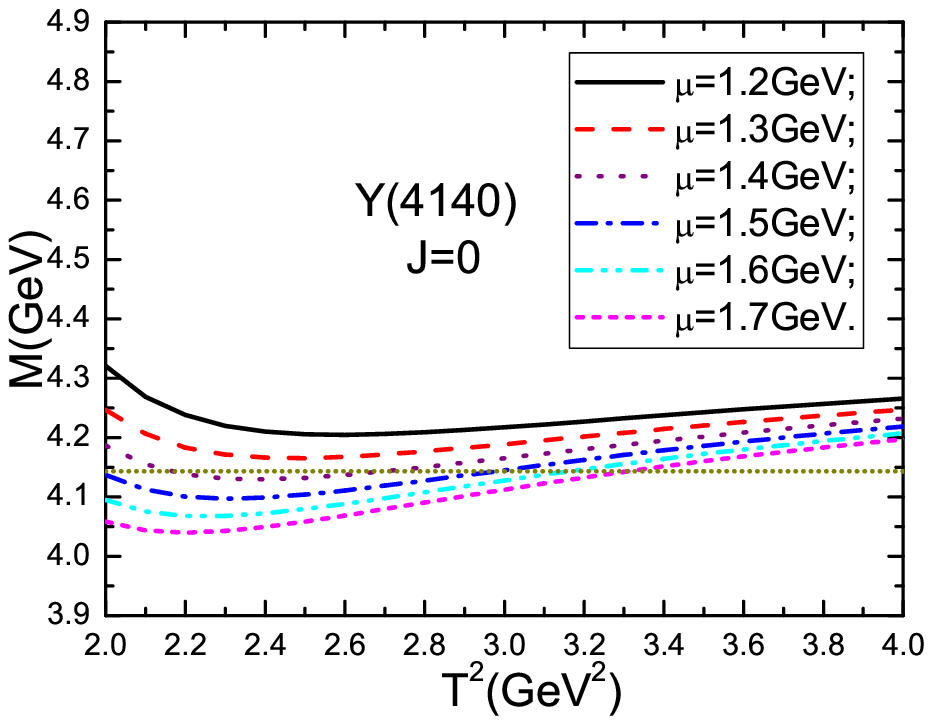}
\includegraphics[totalheight=6cm,width=7cm]{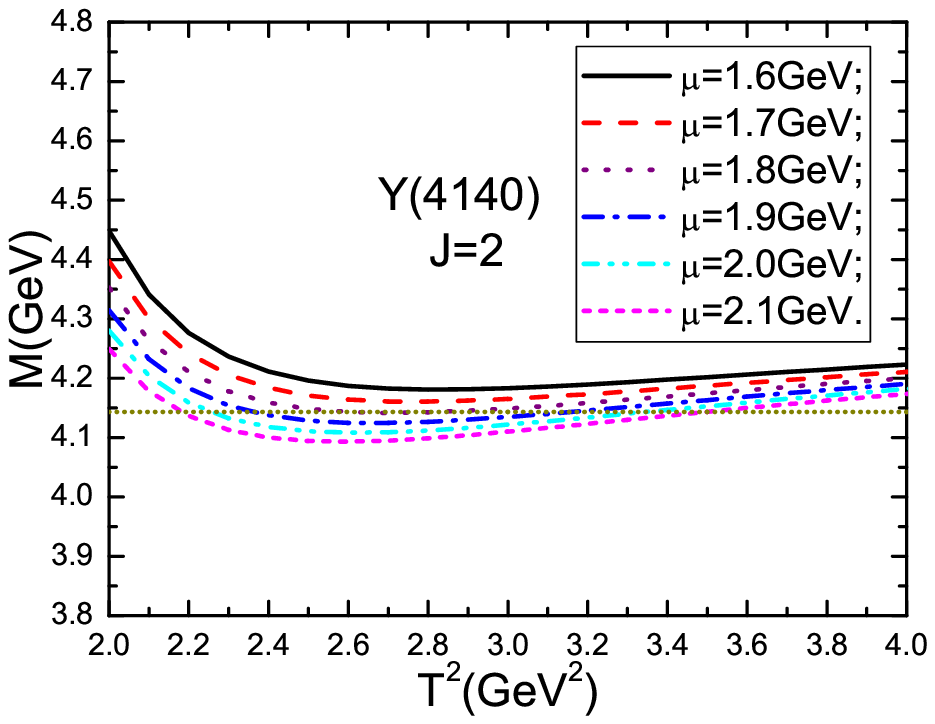}
\includegraphics[totalheight=6cm,width=7cm]{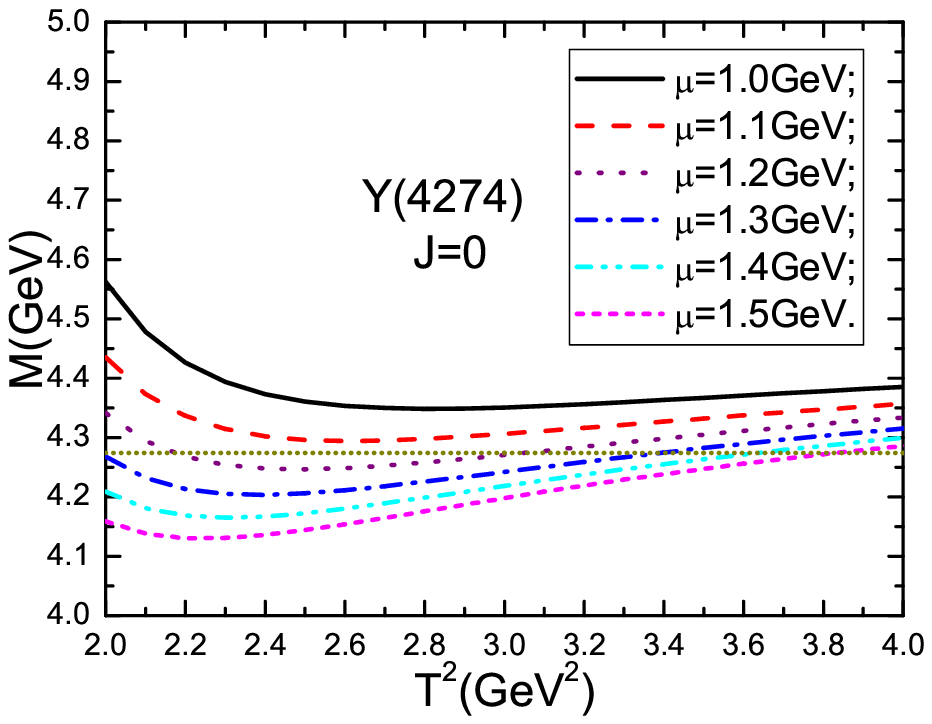}
\includegraphics[totalheight=6cm,width=7cm]{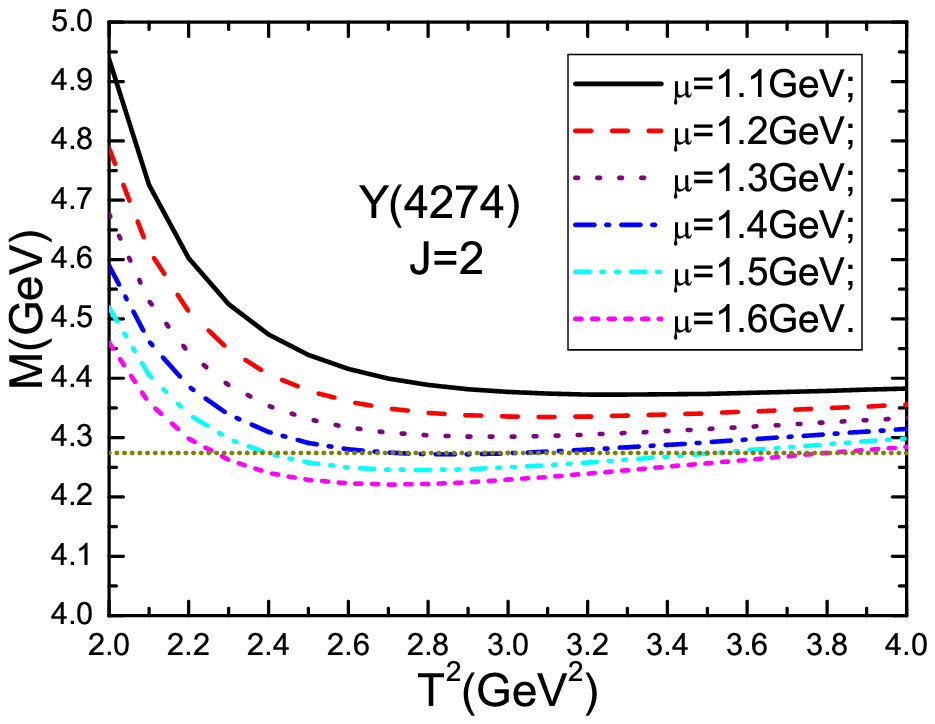}
\includegraphics[totalheight=6cm,width=7cm]{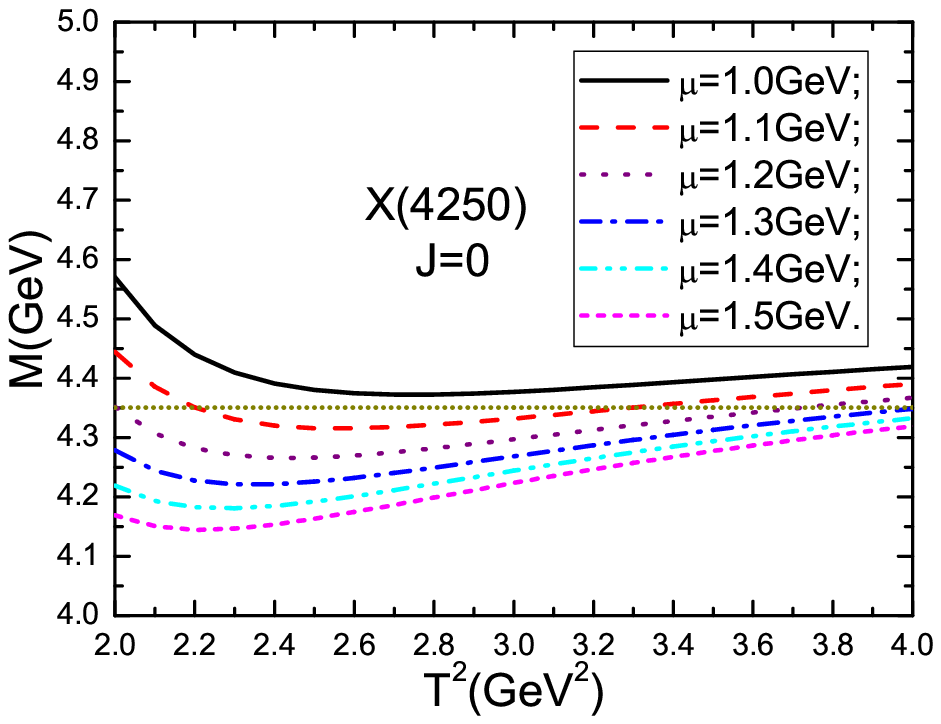}
\includegraphics[totalheight=6cm,width=7cm]{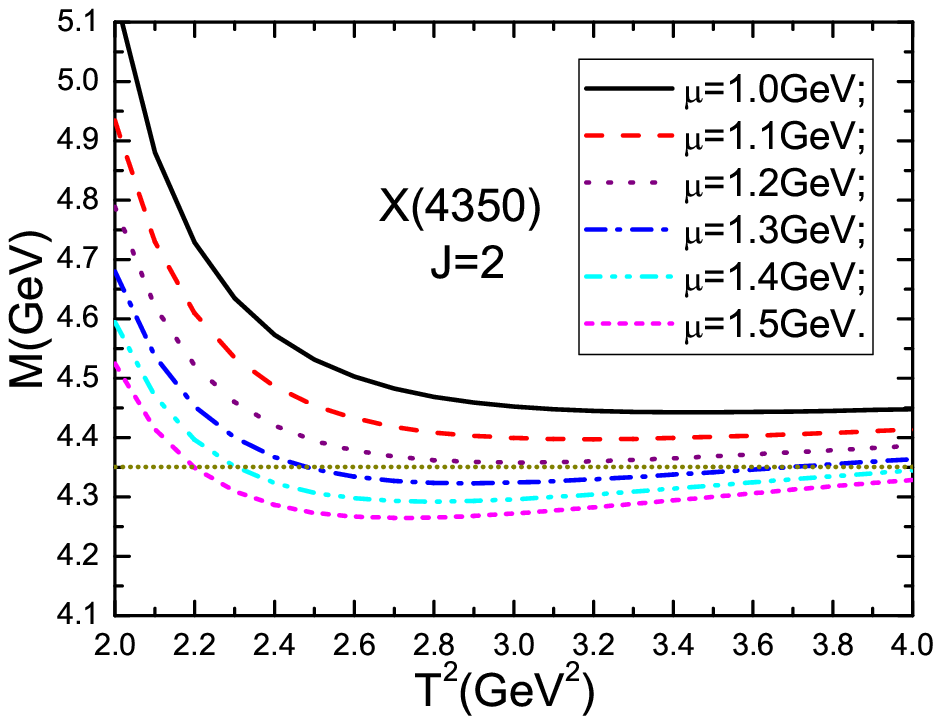}
  \caption{ The masses of the $Y(4140)$, $Y(4274)$ and $X(4350)$ with the assignments $J^{PC}=0^{++}$ and $2^{++}$ respectively  vary with the
  Borel parameters $T^2$ and the energy scales $\mu$, where the horizontal lines denote the experimental values of the masses of the $Y(4140)$, $Y(4274)$ and $X(4350)$, respectively.   }
\end{figure}

\begin{table}
\begin{center}
\begin{tabular}{|c|c|c|c|c|c|}\hline\hline
                  &      $J^{PC}$    & $\sqrt{s_0} (\rm{GeV})$    & $\mu_A(\rm{GeV})$     & $\mu_{T}(\rm{GeV})$   &       \\ \hline
 $Y(4140)$        &      $0^{++}$    & 4.70                       & $1.4-1.7$             & 2.0                   & $\times$  \\   \hline
 $Y(4140)$        &      $2^{++}$    & 4.70                       & $1.8-2.1$             & 2.0                   & $\surd$   \\   \hline   $Y(4274)$        &      $0^{++}$    & 4.80                       & $1.2-1.4$             & 2.3                   & $\times$  \\   \hline
 $Y(4274)$        &      $2^{++}$    & 4.80                       & $1.4-1.6$             & 2.3                   &  $\times$ \\   \hline
 $X(4350)$        &      $0^{++}$    & 4.85                       & $1.1-1.2$             & 2.4                   &  $\times$  \\   \hline
 $X(4350)$        &      $2^{++}$    & 4.85                       & $1.2-1.3$             & 2.4                   &  $\times$  \\   \hline
 \hline
\end{tabular}
\end{center}
\caption{ The continuum threshold parameters $s_0$, allowed energy scales $\mu_A$, theoretical energy scales $\mu_T$ for the $Y(4140)$, $Y(4274)$ and $X(4350)$
with the possible assignments $J^{PC}$, where the $\times$ and $\surd$ denote the compatibility between the $\mu_A$ and $\mu_T$. }
\end{table}

\begin{table}
\begin{center}
\begin{tabular}{|c|c|c|c|c|c|c|c|c|}\hline\hline
$J^{PC}$ &$T^2(\rm{GeV}^2)$ &$\sqrt{s_0} (\rm{GeV})$ &$\mu(\rm{GeV})$ &pole        &$M_{X/Y/Z}(\rm{GeV})$  &$\lambda_{X/Y/Z}$ \\ \hline
$2^{++}$ &$3.0-3.4$         &$4.7\pm0.1$             &2.0             &$(49-69)\%$ &$4.13^{+0.08}_{-0.08}$ &$5.34^{+0.76}_{-0.68}\times10^{-2}\rm{GeV}^5$\\ \hline
$0^{++}$ &$2.5-2.9$         &$4.5\pm0.1$             &1.7             &$(46-70)\%$ &$3.98^{+0.08}_{-0.08}$ &$4.87^{+0.81}_{-0.68}\times10^{-2}\rm{GeV}^5$ \\ \hline
 \hline
\end{tabular}
\end{center}
\caption{ The Borel parameters, continuum threshold parameters, energy scales of the QCD spectral densities, pole contributions, masses and pole residues of the scalar  and tensor  tetraquark states. }
\end{table}

\begin{figure}
\centering
\includegraphics[totalheight=6cm,width=7cm]{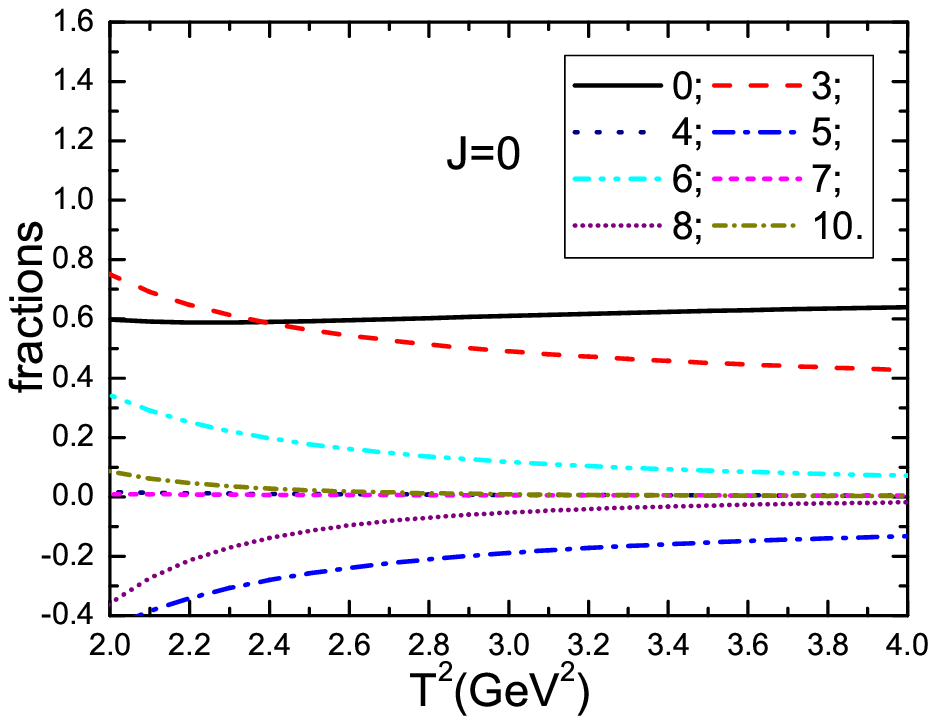}
 \includegraphics[totalheight=6cm,width=7cm]{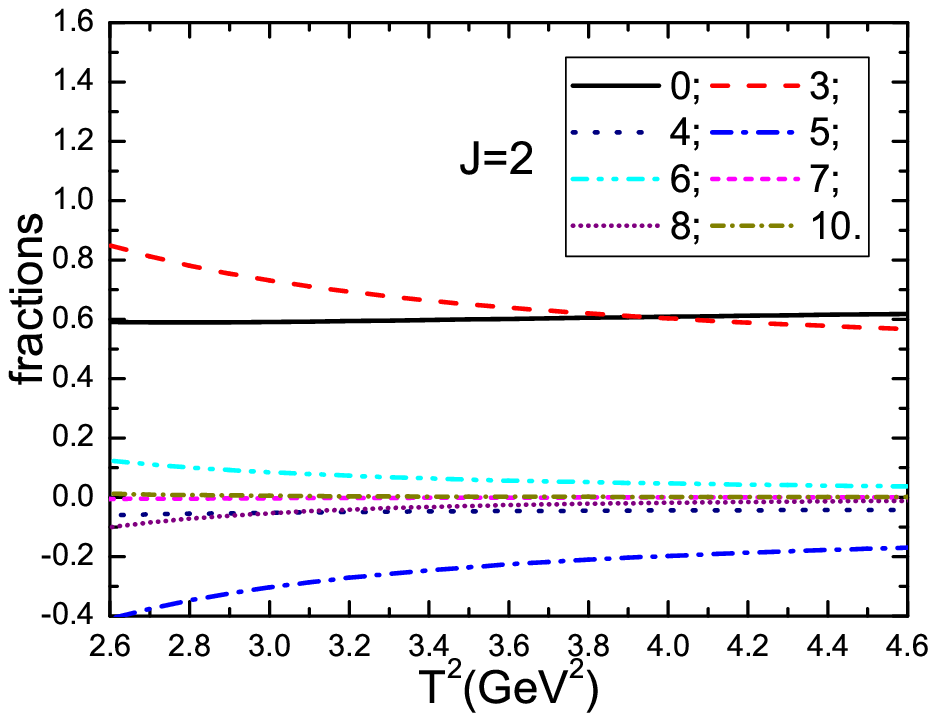}
  \caption{ The contributions of different terms in the operator product expansion for the $J^{PC}=0^{++}$ and $2^{++}$ tetraquark states  with variations of the
  Borel parameters $T^2$, where the 0, 3, 4, 5, 6, 7, 8,  10 denote  the dimensions of the vacuum condensates.   }
\end{figure}

\begin{figure}
\centering
\includegraphics[totalheight=6cm,width=7cm]{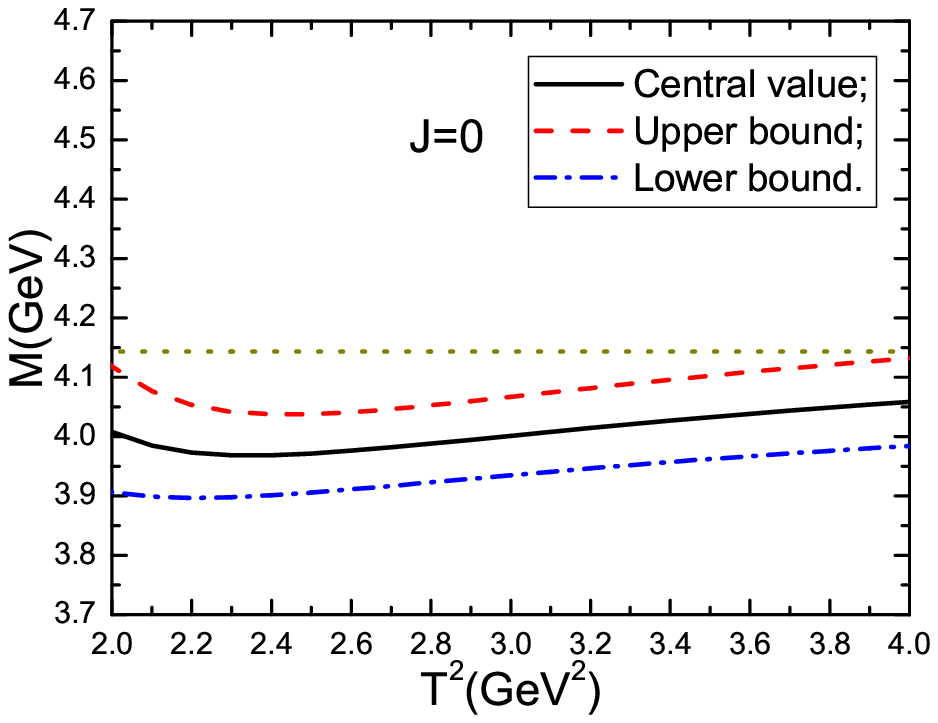}
 \includegraphics[totalheight=6cm,width=7cm]{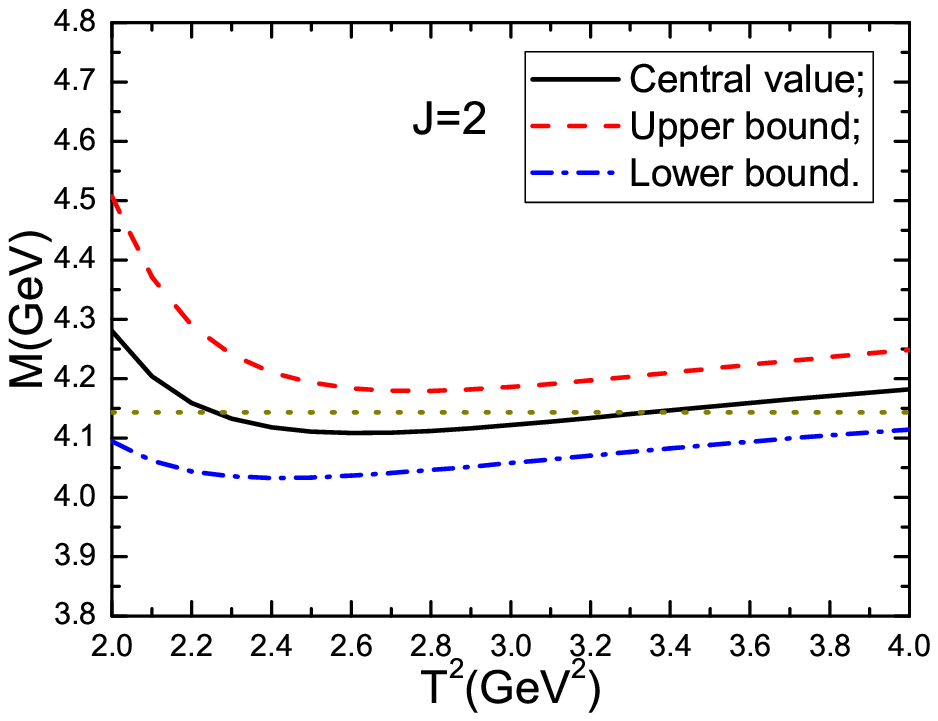}
  \caption{ The masses   of the   $J^{PC}=0^{++}$ and $2^{++}$ tetraquark states  with variations of  the
  Borel parameters $T^2$, where the horizontal lines denote the experimental value of the mass of the $Y(4140)$. }
\end{figure}

\begin{figure}
\centering
\includegraphics[totalheight=6cm,width=7cm]{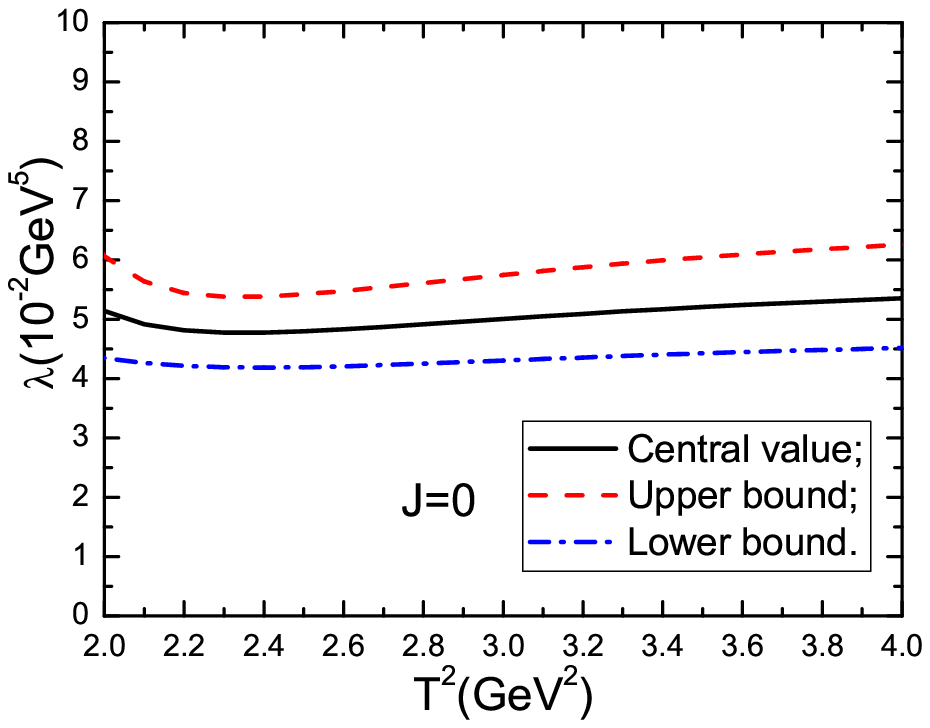}
 \includegraphics[totalheight=6cm,width=7cm]{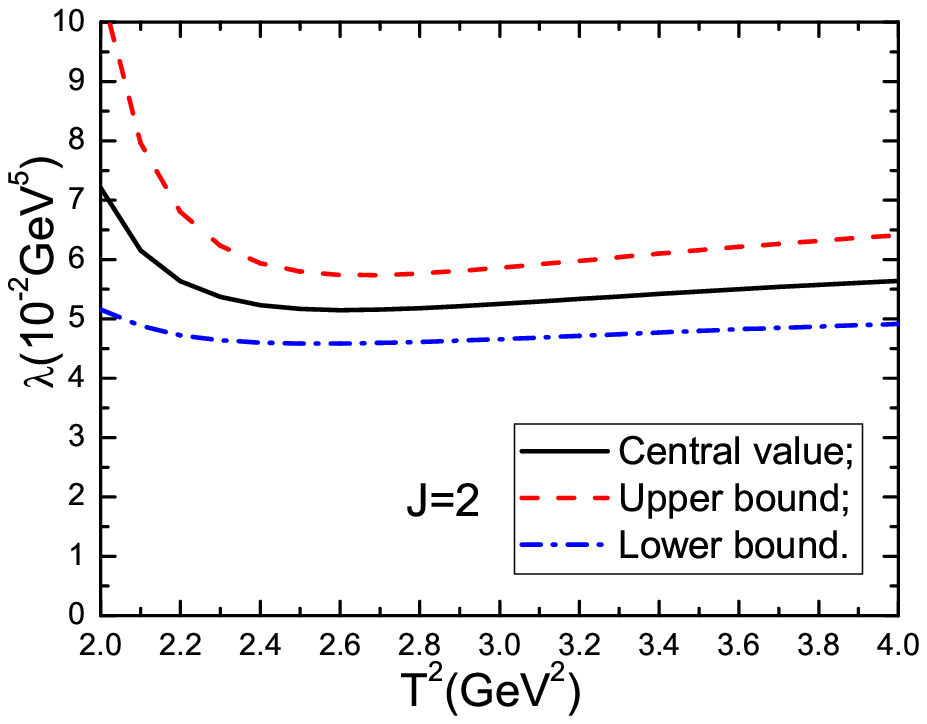}
  \caption{ The pole residues of the   $J^{PC}=0^{++}$ and $2^{++}$ tetraquark states  with variations of  the
  Borel parameters $T^2$.  }
\end{figure}

\section{Conclusion}
In this article, we tentatively assign the $Y(4140)$, $Y(4274)$ and $X(4350)$ to be the scalar and tensor $cs\bar{c}\bar{s}$ tetraquark states, respectively, and study them with the QCD sum rules. In the operator product expansion, we  calculate the contributions of the vacuum condensates up to dimension-10. Furthermore,  we use the  formula $\mu=\sqrt{M^2_{X/Y/Z}-(2{\mathbb{M}}_c)^2}$  to determine  the energy scales of the QCD spectral densities.  The numerical results of the masses $M_{X/Y/Z}$ favor assigning the $Y(4140)$ to be the $J^{PC}=2^{++}$   $cs\bar{c}\bar{s}$ tetraquark state, and disfavor assigning the $Y(4274)$ and $X(4350)$ to be  the $0^{++}$ or $2^{++}$  tetraquark states.  There is no candidate for the scalar $cs\bar{c}\bar{s}$ tetraquark state, we can search for it at the BESIII, LHCb and Belle-II  in the futures.

\section*{Appendix}
The spectral densities $\rho_i(s)$ with $i=0$, 3, 4, 5, 6, 7, 8, 10  at the level of the quark-gluon degrees of
freedom,

\begin{eqnarray}
\rho^{2}_{0}(s)&=&\frac{1}{15360\pi^6}\int_{y_i}^{y_f}dy \int_{z_i}^{1-y}dz \, yz\, (1-y-z)^3\left(s-\overline{m}_c^2\right)^2\left(293s^2-190s\overline{m}_c^2+17\overline{m}_c^4 \right)  \nonumber\\
&&+\frac{1}{5120\pi^6} \int_{y_i}^{y_f}dy \int_{z_i}^{1-y}dz \, yz \,(1-y-z)^2\left(s-\overline{m}_c^2\right)^4    \nonumber\\
&&+\frac{m_sm_c}{128\pi^6}\int_{y_i}^{y_f}dy \int_{z_i}^{1-y}dz \, (y+z)\, (1-y-z)^2 \left(s-\overline{m}_c^2\right)^2\left(4s-\overline{m}_c^2 \right)  \, ,
\end{eqnarray}

\begin{eqnarray}
\rho_{3}^{2}(s)&=&-\frac{m_c\langle \bar{s}s\rangle}{16\pi^4}\int_{y_i}^{y_f}dy \int_{z_i}^{1-y}dz \, (y+z)(1-y-z)\left(s-\overline{m}_c^2\right)\left(3s-\overline{m}_c^2\right)  \nonumber\\
&&+\frac{m_s\langle \bar{s}s\rangle}{160\pi^4}\int_{y_i}^{y_f}dy \int_{z_i}^{1-y}dz \, yz(1-y-z)\left(115s^2-112s\overline{m}_c^2+17\overline{m}_c^4 \right)  \nonumber\\
&&+\frac{m_s\langle \bar{s}s\rangle}{160\pi^4}\int_{y_i}^{y_f}dy \int_{z_i}^{1-y}dz \, yz \left( s - \overline{m}_c^2  \right)^2 \nonumber\\
&&-\frac{m_sm_c^2\langle \bar{s}s\rangle}{4\pi^4}\int_{y_i}^{y_f}dy \int_{z_i}^{1-y}dz   \left( s - \overline{m}_c^2  \right)   \, ,
\end{eqnarray}

\begin{eqnarray}
\rho_{4}^{2}(s)&=&-\frac{m_c^2}{11520\pi^4} \langle\frac{\alpha_s GG}{\pi}\rangle\int_{y_i}^{y_f}dy \int_{z_i}^{1-y}dz \left( \frac{z}{y^2}+\frac{y}{z^2}\right)(1-y-z)^3 \nonumber\\
&&\left\{ 56s-17\overline{m}_c^2+10\overline{m}_c^4\delta\left(s-\overline{m}_c^2\right)\right\} \nonumber\\
&&-\frac{m_c^2}{3840\pi^4}\langle\frac{\alpha_s GG}{\pi}\rangle\int_{y_i}^{y_f}dy \int_{z_i}^{1-y}dz \left(\frac{z}{y^2}+\frac{y}{z^2} \right) (1-y-z)^2 \left(s-\overline{m}_c^2\right) \nonumber\\
&&-\frac{1}{15360\pi^4} \langle\frac{\alpha_s GG}{\pi}\rangle\int_{y_i}^{y_f}dy \int_{z_i}^{1-y}dz \left( y+z\right)(1-y-z)^2 \left( 185s^2-208s\overline{m}_c^2+43\overline{m}_c^4\right) \nonumber\\
&&+\frac{1}{7680\pi^4} \langle\frac{\alpha_s GG}{\pi}\rangle\int_{y_i}^{y_f}dy \int_{z_i}^{1-y}dz \left( y+z\right)(1-y-z) \left( s-\overline{m}_c^2\right)^2 \nonumber\\
&&-\frac{1}{2304\pi^4} \langle\frac{\alpha_s GG}{\pi}\rangle\int_{y_i}^{y_f}dy \int_{z_i}^{1-y}dz \left( y+z\right)(1-y-z)^2 \left( 15s^2-16s\overline{m}_c^2+3\overline{m}_c^4\right) \nonumber\\
&&-\frac{1}{13824\pi^4} \langle\frac{\alpha_s GG}{\pi}\rangle\int_{y_i}^{y_f}dy \int_{z_i}^{1-y}dz \, (1-y-z)^3 \left( 25s^2-24s\overline{m}_c^2+3\overline{m}_c^4\right)  \nonumber\\
&&-\frac{1}{6912\pi^4} \langle\frac{\alpha_s GG}{\pi}\rangle\int_{y_i}^{y_f}dy \int_{z_i}^{1-y}dz  \, yz\,(1-y-z) \left( 25s^2-24s\overline{m}_c^2+3\overline{m}_c^4\right) \nonumber\\
&&-\frac{1}{4608\pi^4} \langle\frac{\alpha_s GG}{\pi}\rangle\int_{y_i}^{y_f}dy \int_{z_i}^{1-y}dz \, (1-y-z)^2 \left( s-\overline{m}_c^2\right)^2 \nonumber\\
&&-\frac{1}{6912\pi^4} \langle\frac{\alpha_s GG}{\pi}\rangle\int_{y_i}^{y_f}dy \int_{z_i}^{1-y}dz \, yz \left( s-\overline{m}_c^2\right)\left( 13s-5\overline{m}_c^2\right) \, ,
\end{eqnarray}

\begin{eqnarray}
\rho^{2}_{5}(s)&=&\frac{m_c\langle \bar{s}g_s\sigma Gs\rangle}{32\pi^4}\int_{y_i}^{y_f}dy \int_{z_i}^{1-y}dz  \, (y+z) \left(2s-\overline{m}_c^2 \right) \nonumber\\
&&+\frac{m_c\langle \bar{s}g_s\sigma Gs\rangle}{144\pi^4}\int_{y_i}^{y_f}dy \int_{z_i}^{1-y}dz  \,  (1-y-z) \left(2s-\overline{m}_c^2 \right)    \nonumber\\
&&-\frac{m_s\langle \bar{s}g_s\sigma Gs\rangle}{480\pi^4}\int_{y_i}^{y_f}dy \int_{z_i}^{1-y}dz \, yz \left\{56s - 17\overline{m}_c^2 +10\overline{m}_c^4 \delta(s-\overline{m}_c^2 )\right\}\nonumber\\
&&-\frac{m_s\langle \bar{s}g_s\sigma Gs\rangle}{480\pi^4}\int_{y_i}^{y_f}dy   \, y(1-y) \left(s - \widetilde{m}_c^2 \right)+\frac{m_sm_c^2\langle \bar{s}g_s\sigma Gs\rangle}{16\pi^4}\int_{y_i}^{y_f}dy  \nonumber\\
&&+\frac{m_sm_c^2\langle \bar{s}g_s\sigma Gs\rangle}{288\pi^4}\int_{y_i}^{y_f}dy \int_{z_i}^{1-y}dz  \left(\frac{1}{y}+\frac{1}{z} \right) \, ,
\end{eqnarray}

\begin{eqnarray}
\rho_{6}^{2}(s)&=&\frac{m_c^2\langle\bar{s}s\rangle^2}{6\pi^2}\int_{y_i}^{y_f}dy   +\frac{g_s^2\langle\bar{s}s\rangle^2}{3240\pi^4}\int_{y_i}^{y_f}dy \int_{z_i}^{1-y}dz\, yz \left\{56s-17\overline{m}_c^2 +10\overline{m}_c^4\delta\left(s-\overline{m}_c^2 \right)\right\}\nonumber\\
&&+\frac{g_s^2\langle\bar{s}s\rangle^2}{3240\pi^4}\int_{y_i}^{y_f}dy \,y(1-y)\left(s-\widetilde{m}_c^2 \right)  \nonumber\\
&&-\frac{g_s^2\langle\bar{s}s\rangle^2}{9720\pi^4}\int_{y_i}^{y_f}dy \int_{z_i}^{1-y}dz \, (1-y-z)\left\{ 45\left(\frac{z}{y}+\frac{y}{z} \right)\left(2s-\overline{m}_c^2 \right)+\left(\frac{z}{y^2}+\frac{y}{z^2} \right)\right.\nonumber\\
&&\left.m_c^2\left[ 19+20\overline{m}_c^2\delta\left(s-\overline{m}_c^2 \right)\right]+(y+z)\left[18\left(3s-\overline{m}_c^2\right)+10\overline{m}_c^4\delta\left(s-\overline{m}_c^2\right) \right] \right\} \nonumber\\
&&-\frac{g_s^2\langle\bar{s}s\rangle^2}{9720\pi^4}\int_{y_i}^{y_f}dy \int_{z_i}^{1-y}dz \, (1-y-z)\left\{  15\left(\frac{z}{y}+\frac{y}{z} \right)\left(2s-\overline{m}_c^2 \right)+\left(\frac{z}{y^2}+\frac{y}{z^2} \right)\right. \nonumber\\
&&\left.m_c^2\left[ 6+5\overline{m}_c^2\delta\left(s-\overline{m}_c^2\right)\right]+(y+z)\left[56s-17\overline{m}_c^2 +10\overline{m}_c^4\delta\left(s-\overline{m}_c^2\right)\right] \right\}\nonumber\\
&&-\frac{m_sm_c \langle\bar{s}s\rangle^2}{12\pi^2}\int_{y_i}^{y_f}dy \left\{ 1+\widetilde{m}_c^2\delta(s-\widetilde{m}_c^2)\right\}\, ,
\end{eqnarray}

\begin{eqnarray}
\rho_7^{2}(s)&=&\frac{m_c^3\langle\bar{s}s\rangle}{144\pi^2 T^2}\langle\frac{\alpha_sGG}{\pi}\rangle\int_{y_i}^{y_f}dy \int_{z_i}^{1-y}dz \left(\frac{y}{z^3}+\frac{z}{y^3}+\frac{1}{y^2}+\frac{1}{z^2}\right)(1-y-z)\, \overline{m}_c^2 \, \delta\left(s-\overline{m}_c^2\right)\nonumber\\
&&-\frac{m_c\langle\bar{s}s\rangle}{48\pi^2}\langle\frac{\alpha_sGG}{\pi}\rangle\int_{y_i}^{y_f}dy \int_{z_i}^{1-y}dz \left(\frac{y}{z^2}+\frac{z}{y^2}\right)(1-y-z)  \left\{1+\overline{m}_c^2\delta\left(s-\overline{m}_c^2\right) \right\}\nonumber\\
&&+\frac{m_c\langle\bar{s}s\rangle}{48\pi^2}\langle\frac{\alpha_sGG}{\pi}\rangle\int_{y_i}^{y_f}dy \int_{z_i}^{1-y}dz\left\{1+\frac{\overline{m}_c^2}{3}\delta\left(s-\overline{m}_c^2\right) \right\} \nonumber\\
&&+\frac{m_c\langle\bar{s}s\rangle}{432\pi^2}\langle\frac{\alpha_sGG}{\pi}\rangle\int_{y_i}^{y_f}dy \int_{z_i}^{1-y}dz\left(\frac{1-y}{y}+\frac{1-z}{z}\right)
\left\{1+\overline{m}_c^2\delta\left(s-\overline{m}_c^2\right) \right\}\nonumber \\
&&-\frac{m_c\langle\bar{s}s\rangle}{288\pi^2}\langle\frac{\alpha_sGG}{\pi}\rangle\int_{y_i}^{y_f}dy \left\{1+ \widetilde{m}_c^2 \, \delta \left(s-\widetilde{m}_c^2\right) \right\}\, ,
\end{eqnarray}

\begin{eqnarray}
\rho_8^{2}(s)&=&-\frac{m_c^2\langle\bar{s}s\rangle\langle\bar{s}g_s\sigma Gs\rangle}{12\pi^2}\int_0^1 dy \left(1+\frac{\widetilde{m}_c^2}{T^2} \right)\delta\left(s-\widetilde{m}_c^2\right)\nonumber \\
&&-\frac{ m_c^2\langle\bar{s}s\rangle\langle\bar{s}g_s\sigma Gs\rangle}{216\pi^2}\int_{0}^{1} dy \frac{1}{y(1-y)}\delta\left(s-\widetilde{m}_c^2\right)
 \, ,
\end{eqnarray}

\begin{eqnarray}
\rho_{10}^{2}(s)&=&\frac{m_c^2\langle\bar{s}g_s\sigma Gs\rangle^2}{96\pi^2T^6}\int_0^1 dy \, \widetilde{m}_c^4 \, \delta \left( s-\widetilde{m}_c^2\right)
\nonumber \\
&&-\frac{m_c^4\langle\bar{s}s\rangle^2}{108T^4}\langle\frac{\alpha_sGG}{\pi}\rangle\int_0^1 dy  \left\{ \frac{1}{y^3}+\frac{1}{(1-y)^3}\right\} \delta\left( s-\widetilde{m}_c^2\right)\nonumber\\
&&+\frac{m_c^2\langle\bar{s}s\rangle^2}{36T^2}\langle\frac{\alpha_sGG}{\pi}\rangle\int_0^1 dy  \left\{ \frac{1}{y^2}+\frac{1}{(1-y)^2}\right\} \delta\left( s-\widetilde{m}_c^2\right)\nonumber\\
&&-\frac{m_c^2\langle\bar{s}s\rangle^2}{324T^2}\langle\frac{\alpha_sGG}{\pi}\rangle\int_0^1 dy   \frac{1}{y(1-y)} \delta\left( s-\widetilde{m}_c^2\right)\nonumber \\
&&+\frac{m_c^2\langle\bar{s}g_s\sigma Gs\rangle^2}{864 \pi^2T^4} \int_0^1 dy   \frac{1}{y(1-y)}  \widetilde{m}_c^2 \, \delta\left( s-\widetilde{m}_c^2\right)\nonumber\\
&&+\frac{m_c^2\langle\bar{s}g_s\sigma Gs\rangle^2}{576 \pi^2T^2} \int_0^1 dy   \frac{1}{y(1-y)}   \delta\left( s-\widetilde{m}_c^2\right)\nonumber \\
&&+\frac{m_c^2\langle\bar{s} s\rangle^2}{108 T^6}\langle\frac{\alpha_sGG}{\pi}\rangle\int_0^1 dy \, \widetilde{m}_c^4 \, \delta \left( s-\widetilde{m}_c^2\right) \, ,
\end{eqnarray}

\begin{eqnarray}
\rho^{0}_{0}(s)&=&\frac{1}{256\pi^6}\int_{y_i}^{y_f}dy \int_{z_i}^{1-y}dz \, yz\, (1-y-z)^3\left(s-\overline{m}_c^2\right)^2\left(7s^2-6s\overline{m}_c^2+\overline{m}_c^4 \right)  \nonumber\\
&&+\frac{1}{256\pi^6} \int_{y_i}^{y_f}dy \int_{z_i}^{1-y}dz \, yz \,(1-y-z)^2\left(s-\overline{m}_c^2\right)^3 \left(3s-\overline{m}_c^2\right)  \nonumber\\
&&+\frac{m_sm_c}{128\pi^6}\int_{y_i}^{y_f}dy \int_{z_i}^{1-y}dz \, (y+z)\, (1-y-z)^2 \left(s-\overline{m}_c^2\right)^2\left(5s-2\overline{m}_c^2 \right)  \, ,
\end{eqnarray}

\begin{eqnarray}
\rho_{3}^{0}(s)&=&-\frac{m_c\langle \bar{s}s\rangle}{8\pi^4}\int_{y_i}^{y_f}dy \int_{z_i}^{1-y}dz \, (y+z)(1-y-z)\left(s-\overline{m}_c^2\right)\left(2s-\overline{m}_c^2\right)  \nonumber\\
&&+\frac{m_s\langle \bar{s}s\rangle}{8\pi^4}\int_{y_i}^{y_f}dy \int_{z_i}^{1-y}dz \, yz(1-y-z)\left(10s^2-12s\overline{m}_c^2+3\overline{m}_c^4 \right)   \nonumber\\
&&+\frac{m_s\langle \bar{s}s\rangle}{8\pi^4}\int_{y_i}^{y_f}dy \int_{z_i}^{1-y}dz \, yz\left(s-\overline{m}_c^2\right)\left(2s-\overline{m}_c^2\right)  \nonumber\\
&&-\frac{m_sm_c^2\langle \bar{s}s\rangle}{2\pi^4}\int_{y_i}^{y_f}dy \int_{z_i}^{1-y}dz  \left(s-\overline{m}_c^2\right)   \, ,
\end{eqnarray}

\begin{eqnarray}
\rho_{4}^{0}(s)&=&-\frac{m_c^2}{192\pi^4} \langle\frac{\alpha_s GG}{\pi}\rangle\int_{y_i}^{y_f}dy \int_{z_i}^{1-y}dz \left( \frac{z}{y^2}+\frac{y}{z^2}\right)(1-y-z)^3 \nonumber\\
&&\left\{ 2s-\overline{m}_c^2+\frac{\overline{m}_c^4}{6}\delta\left(s-\overline{m}_c^2\right)\right\} \nonumber\\
&&-\frac{m_c^2}{384\pi^4}\langle\frac{\alpha_s GG}{\pi}\rangle\int_{y_i}^{y_f}dy \int_{z_i}^{1-y}dz \left(\frac{z}{y^2}+\frac{y}{z^2} \right) (1-y-z)^2 \left(3s-2\overline{m}_c^2\right) \nonumber\\
&&-\frac{1}{768\pi^4} \langle\frac{\alpha_s GG}{\pi}\rangle\int_{y_i}^{y_f}dy \int_{z_i}^{1-y}dz \left( y+z\right)(1-y-z)^2 \left( 10s^2-12s\overline{m}_c^2+3\overline{m}_c^4\right) \nonumber\\
&&+\frac{1}{384\pi^4} \langle\frac{\alpha_s GG}{\pi}\rangle\int_{y_i}^{y_f}dy \int_{z_i}^{1-y}dz \left( y+z\right)(1-y-z) \left( s-\overline{m}_c^2\right)\left( 2s-\overline{m}_c^2\right) \nonumber\\
&&+\frac{1}{384\pi^4} \langle\frac{\alpha_s GG}{\pi}\rangle\int_{y_i}^{y_f}dy \int_{z_i}^{1-y}dz \left( y+z\right)(1-y-z)^2 \left( 10s^2-12s\overline{m}_c^2+3\overline{m}_c^4\right) \nonumber\\
&&+\frac{1}{3456\pi^4} \langle\frac{\alpha_s GG}{\pi}\rangle\int_{y_i}^{y_f}dy \int_{z_i}^{1-y}dz \, (1-y-z)^3 \left( 10s^2-12s\overline{m}_c^2+3\overline{m}_c^4\right)  \nonumber\\
&&+\frac{1}{576\pi^4} \langle\frac{\alpha_s GG}{\pi}\rangle\int_{y_i}^{y_f}dy \int_{z_i}^{1-y}dz  \, yz\,(1-y-z) \left( 10s^2-12s\overline{m}_c^2+3\overline{m}_c^4\right) \nonumber\\
&&+\frac{1}{576\pi^4} \langle\frac{\alpha_s GG}{\pi}\rangle\int_{y_i}^{y_f}dy \int_{z_i}^{1-y}dz \, (1-y-z)^2 \left( s-\overline{m}_c^2\right)
\left( 2s-\overline{m}_c^2\right) \nonumber\\
&&+\frac{1}{288\pi^4} \langle\frac{\alpha_s GG}{\pi}\rangle\int_{y_i}^{y_f}dy \int_{z_i}^{1-y}dz \, yz \left( s-\overline{m}_c^2\right)\left( 2s-\overline{m}_c^2\right) \, ,
\end{eqnarray}

\begin{eqnarray}
\rho^{0}_{5}(s)&=&\frac{m_c\langle \bar{s}g_s\sigma Gs\rangle}{32\pi^4}\int_{y_i}^{y_f}dy \int_{z_i}^{1-y}dz  \, (y+z) \left(3s-2\overline{m}_c^2 \right) \nonumber\\
&&-\frac{m_c\langle \bar{s}g_s\sigma Gs\rangle}{48\pi^4}\int_{y_i}^{y_f}dy \int_{z_i}^{1-y}dz  \,  (1-y-z) \left(3s-2\overline{m}_c^2 \right)     \nonumber\\
&&-\frac{m_s\langle \bar{s}g_s\sigma Gs\rangle}{8\pi^4}\int_{y_i}^{y_f}dy \int_{z_i}^{1-y}dz  \,  yz\left\{2s-\overline{m}_c^2+\frac{\overline{m}_c^2}{6} \delta\left(s-\overline{m}_c^2 \right) \right\}    \nonumber\\
&&-\frac{m_s\langle \bar{s}g_s\sigma Gs\rangle}{48\pi^4}\int_{y_i}^{y_f}dy    \,  y(1-y)\left(3s-2\widetilde{m}_c^2 \right)     \nonumber\\
&&+\frac{m_sm_c^2\langle \bar{s}g_s\sigma Gs\rangle}{8\pi^4}\int_{y_i}^{y_f}dy        \nonumber\\
&&-\frac{m_sm_c^2\langle \bar{s}g_s\sigma Gs\rangle}{48\pi^4}\int_{y_i}^{y_f}dy \int_{z_i}^{1-y}dz  \left(\frac{1}{y}+\frac{1}{z} \right) \, ,
\end{eqnarray}

\begin{eqnarray}
\rho_{6}^{0}(s)&=&\frac{m_c^2\langle\bar{s}s\rangle^2}{3\pi^2}\int_{y_i}^{y_f}dy   +\frac{g_s^2\langle\bar{s}s\rangle^2}{54\pi^4}\int_{y_i}^{y_f}dy \int_{z_i}^{1-y}dz\, yz \left\{2s-\overline{m}_c^2 +\frac{\overline{m}_c^4}{6}\delta\left(s-\overline{m}_c^2 \right)\right\}\nonumber\\
&&+\frac{g_s^2\langle\bar{s}s\rangle^2}{324\pi^4}\int_{y_i}^{y_f}dy \,y(1-y)\left(3s-2\widetilde{m}_c^2 \right)  \nonumber\\
&&-\frac{g_s^2\langle\bar{s}s\rangle^2}{648\pi^4}\int_{y_i}^{y_f}dy \int_{z_i}^{1-y}dz \, (1-y-z)\left\{ 3\left(\frac{z}{y}+\frac{y}{z} \right)\left(3s-2\overline{m}_c^2 \right)+\left(\frac{z}{y^2}+\frac{y}{z^2} \right)\right.\nonumber\\
&&\left.m_c^2\left[ 2+ \overline{m}_c^2\delta\left(s-\overline{m}_c^2 \right)\right]+(y+z)\left[12\left(2s-\overline{m}_c^2\right)+2\overline{m}_c^4\delta\left(s-\overline{m}_c^2\right) \right] \right\} \nonumber\\
&&-\frac{g_s^2\langle\bar{s}s\rangle^2}{1944\pi^4}\int_{y_i}^{y_f}dy \int_{z_i}^{1-y}dz \, (1-y-z)\left\{  15\left(\frac{z}{y}+\frac{y}{z} \right)\left(3s-2\overline{m}_c^2 \right)+7\left(\frac{z}{y^2}+\frac{y}{z^2} \right)\right. \nonumber\\
&&\left.m_c^2\left[ 2+\overline{m}_c^2\delta\left(s-\overline{m}_c^2\right)\right]+(y+z)\left[12\left(2s-\overline{m}_c^2\right) +2\overline{m}_c^4\delta\left(s-\overline{m}_c^2\right)\right] \right\} \nonumber\\
&&-\frac{m_sm_c \langle\bar{s}s\rangle^2}{12\pi^2}\int_{y_i}^{y_f}dy \left\{ 2+\widetilde{m}_c^2\delta(s-\widetilde{m}_c^2)\right\}\, ,
\end{eqnarray}

\begin{eqnarray}
\rho_7^{0}(s)&=&\frac{m_c^3\langle\bar{s}s\rangle}{144\pi^2  }\langle\frac{\alpha_sGG}{\pi}\rangle\int_{y_i}^{y_f}dy \int_{z_i}^{1-y}dz \left(\frac{y}{z^3}+\frac{z}{y^3}+\frac{1}{y^2}+\frac{1}{z^2}\right)(1-y-z)\nonumber\\
&&\left(1+\frac{ \overline{m}_c^2}{T^2}\right) \delta\left(s-\overline{m}_c^2\right)\nonumber\\
&&-\frac{m_c\langle\bar{s}s\rangle}{48\pi^2}\langle\frac{\alpha_sGG}{\pi}\rangle\int_{y_i}^{y_f}dy \int_{z_i}^{1-y}dz \left(\frac{y}{z^2}+\frac{z}{y^2}\right)(1-y-z)  \left\{2+\overline{m}_c^2\delta\left(s-\overline{m}_c^2\right) \right\}\nonumber\\
&&+\frac{m_c\langle\bar{s}s\rangle}{48\pi^2}\langle\frac{\alpha_sGG}{\pi}\rangle\int_{y_i}^{y_f}dy \int_{z_i}^{1-y}dz\left\{2+ \overline{m}_c^2 \delta\left(s-\overline{m}_c^2\right) \right\} \nonumber\\
&&-\frac{m_c\langle\bar{s}s\rangle}{144\pi^2}\langle\frac{\alpha_sGG}{\pi}\rangle\int_{y_i}^{y_f}dy \int_{z_i}^{1-y}dz\left(\frac{1-y}{y}+\frac{1-z}{z}\right)
\left\{2+\overline{m}_c^2\delta\left(s-\overline{m}_c^2\right) \right\}\nonumber \\
&&-\frac{m_c\langle\bar{s}s\rangle}{288\pi^2}\langle\frac{\alpha_sGG}{\pi}\rangle\int_{y_i}^{y_f}dy \left\{2+ \widetilde{m}_c^2 \, \delta \left(s-\widetilde{m}_c^2\right) \right\}\, ,
\end{eqnarray}

\begin{eqnarray}
\rho_8^{0}(s)&=&-\frac{m_c^2\langle\bar{s}s\rangle\langle\bar{s}g_s\sigma Gs\rangle}{6\pi^2}\int_0^1 dy \left(1+\frac{\widetilde{m}_c^2}{T^2} \right)\delta\left(s-\widetilde{m}_c^2\right)\nonumber \\
&&+\frac{ m_c^2\langle\bar{s}s\rangle\langle\bar{s}g_s\sigma Gs\rangle}{36\pi^2}\int_{0}^{1} dy \frac{1}{y(1-y)}\delta\left(s-\widetilde{m}_c^2\right)
 \, ,
\end{eqnarray}

\begin{eqnarray}
\rho_{10}^{0}(s)&=&\frac{m_c^2\langle\bar{s}g_s\sigma Gs\rangle^2}{48\pi^2T^6}\int_0^1 dy \, \widetilde{m}_c^4 \, \delta \left( s-\widetilde{m}_c^2\right)
\nonumber \\
&&-\frac{m_c^4\langle\bar{s}s\rangle^2}{54T^4}\langle\frac{\alpha_sGG}{\pi}\rangle\int_0^1 dy  \left\{ \frac{1}{y^3}+\frac{1}{(1-y)^3}\right\} \delta\left( s-\widetilde{m}_c^2\right)\nonumber\\
&&+\frac{m_c^2\langle\bar{s}s\rangle^2}{18T^2}\langle\frac{\alpha_sGG}{\pi}\rangle\int_0^1 dy  \left\{ \frac{1}{y^2}+\frac{1}{(1-y)^2}\right\} \delta\left( s-\widetilde{m}_c^2\right)\nonumber\\
&&+\frac{m_c^2\langle\bar{s}s\rangle^2}{54T^2}\langle\frac{\alpha_sGG}{\pi}\rangle\int_0^1 dy   \frac{1}{y(1-y)} \delta\left( s-\widetilde{m}_c^2\right)\nonumber \\
&&-\frac{m_c^2\langle\bar{s}g_s\sigma Gs\rangle^2}{144 \pi^2T^4} \int_0^1 dy   \frac{1}{y(1-y)}  \widetilde{m}_c^2 \, \delta\left( s-\widetilde{m}_c^2\right)\nonumber\\
&&+\frac{m_c^2\langle\bar{s}g_s\sigma Gs\rangle^2}{32 \pi^2T^2} \int_0^1 dy   \frac{1}{y(1-y)}   \delta\left( s-\widetilde{m}_c^2\right)\nonumber \\
&&+\frac{m_c^2\langle\bar{s} s\rangle^2}{54 T^6}\langle\frac{\alpha_sGG}{\pi}\rangle\int_0^1 dy \, \widetilde{m}_c^4 \, \delta \left( s-\widetilde{m}_c^2\right) \, ,
\end{eqnarray}
 the subscripts $0$, $3$, $4$, $5$, $6$, $7$, $8$ and $10$ denote the dimensions of the vacuum condensates, the superscripts $0$ and $2$ denote the spin the tetraquark states, the $T^2$ denotes the Borel parameter; $y_{f}=\frac{1+\sqrt{1-4m_c^2/s}}{2}$,
$y_{i}=\frac{1-\sqrt{1-4m_c^2/s}}{2}$, $z_{i}=\frac{y
m_c^2}{y s -m_c^2}$, $\overline{m}_c^2=\frac{(y+z)m_c^2}{yz}$,
$ \widetilde{m}_c^2=\frac{m_c^2}{y(1-y)}$, $\int_{y_i}^{y_f}dy \to \int_{0}^{1}dy$, $\int_{z_i}^{1-y}dz \to \int_{0}^{1-y}dz$,
 when the $\delta$ functions $\delta\left(s-\overline{m}_c^2\right)$ and $\delta\left(s-\widetilde{m}_c^2\right)$ appear.

\section*{Acknowledgements}
This  work is supported by National Natural Science Foundation,
Grant Numbers 11375063,  and Natural Science Foundation of Hebei province, Grant Number A2014502017.

\end{document}